\documentclass[twocolumn,prl,amsmath,amssymb,showpacs]{revtex4-1}
\usepackage{graphicx,color}
\usepackage{amsmath,amssymb,latexsym,amsthm}
\usepackage{pgfplots}
\usepackage{mathtools}
\usepackage{mhchem}

\usepackage{tikz}
\usetikzlibrary{chains,shapes,fit,calc,arrows}
\usepackage{pgfplots, pgfplotstable}
\usepgfplotslibrary{groupplots}

\usepackage{color}

\newcommand{\yp}{y_{\mathrm p}}
\newcommand{\muRef}{\mu^{\mathrm{ref}}_{y_{\mathrm p}}}
\newcommand{\de}{\mathrm d}
\newcommand{\dt}{{\mathrm d}_t}
\renewcommand{\vec}[1]{\boldsymbol{#1}}

\newcommand{\caF}{{\mathcal F}}
\newcommand{\caG}{{\mathcal G}}
\newcommand{\caL}{{\mathcal L}}

\newcommand{\caV}{{\mathcal V}}
\newcommand{\caJ}{{\mathcal J}}

\definecolor{webgreen}{rgb}{0,.5,0}
\definecolor{webbrown}{rgb}{.6,0,0}
\definecolor{grigio}{rgb}{.85,.85,.85} 
\definecolor{RoyalBlue}{rgb}{0.0, 0.14, 0.4}
\definecolor{skyblue1}{rgb}{0.45,0.62,0.81}
\definecolor{skyblue2}{rgb}{0.2,0.39,0.64}
\definecolor{skyblue3}{rgb}{0.13,0.29,0.53}
\definecolor{scarlet1}{rgb}{0.93,0.16,0.16}
\definecolor{scarlet2}{rgb}{0.8,0,0}
\definecolor{scarlet3}{rgb}{0.64,0,0}

\usepackage{hyperref}
\hypersetup{%
    colorlinks=true, linktocpage=true, pdfstartpage=1, pdfstartview=FitV,%
    breaklinks=true, pdfpagemode=UseNone, pageanchor=true, pdfpagemode=UseOutlines,%
    plainpages=false, bookmarksnumbered, bookmarksopen=true, bookmarksopenlevel=1,%
    hypertexnames=true, pdfhighlight=/O,
    urlcolor=webbrown, linkcolor=RoyalBlue, citecolor=webgreen, 
    pdftitle={Information Thermodynamics of Turing Patterns},%
    pdfauthor={Gianmaria Falasco},%
    pdfsubject={},%
    pdfkeywords={},%
    pdfcreator={pdfLaTeX},%
    pdfproducer={LaTeX REVTeX}%
}

\begin{document}

\title{Information Thermodynamics of Turing Patterns}
\newcommand\unilu{\affiliation{Complex Systems and Statistical Mechanics, Physics and Materials Science Research Unit, University of Luxembourg, L-1511 Luxembourg}}
\author{Gianmaria Falasco}
\unilu
\author{Riccardo Rao}
\unilu
\author{Massimiliano Esposito}
\unilu

\date{\today}

\begin{abstract}

We set up a rigorous thermodynamic description of reaction-diffusion systems driven out of equilibrium by  time-dependent space-distributed chemostats. Building on the assumption of local equilibrium, nonequilibrium thermodynamic potentials are constructed exploiting the symmetries of the chemical network topology. It is shown that the canonical (resp. semigrand canonical) nonequilibrium free energy works as a Lyapunov function in the relaxation to equilibrium of a closed (resp. open) system and its variation provides the minimum amount of work needed to manipulate the species concentrations. The theory is used to study  analytically the Turing pattern formation in a prototypical reaction-diffusion system, the one-dimensional Brusselator model, and to classify it as a genuine thermodynamic nonequilibrium phase transition.

\end{abstract}

\pacs{05.70.Ln, 87.16.Yc}

\maketitle

\emph{Introduction---}Reaction-diffusion systems (RDS) are ubiquitous in nature.
When nonlinear feedback effects within the chemical reactions are locally destabilized by diffusion, complex spatiotemporal phenomena emerge. These latter, ranging from stationary Turing patterns \cite{castets90, ouyang91} to travelling waves \cite{zaikin70, winfree72}, play a critical role in the aggregation and structuring of hard matter \cite{ortolev94} as well as living systems \cite{murray01}. In biology, striking examples are embryogenesis determined by the pre-patterning of morphogens \cite{kondo10, kretschmer16} and cellular rhythms regulated by calcium waves \cite{falke04, thurley12}. 

Nonequilibrium conditions, consisting in a continual influx of chemicals and energy, are required to create and maintain these dissipative structures.
Since the original work of Prigogine \cite{prigogine71, nicolis77}, which made clear how order can emerge spontaneously at the expense of continuous dissipation, much work has been dedicated to better understanding the chaotic and nonequilibrium dynamics of RDS \cite{cross93}.
Most of it has focused on searching for general extremum principles, e.g. in selecting the relative stability of competing patterns \cite{*[][{, ch.~5.}] {ross08}}.
Nevertheless, a complete framework is still lacking that models RDS as proper  thermodynamic systems, in contact with nonequilibrium chemical reservoirs, subject to external work and entropy changes.
Such a theory is all the more necessary nowadays, when promising technological applications, such as biomimetics \cite{rossi08, grzybowski16} and chemical computing \cite{adamatzky05}, are envisaged that deliberately exploit the self-organized structures of RDS. 
In this respect, the work needed to manipulate a Turing pattern and the efficiency with which information exchanges through travelling waves can occur are thermodynamic questions of crucial importance.

In this Letter we present a rigorous thermodynamic theory of RDS far from equilibrium.
We take the viewpoint of stochastic thermodynamics \cite{jarzynski11,seifert12,vandenbroeck15} and carry over its systematic way to define thermodynamic quantities (such as work and entropy), anchoring them to the (herein deterministic) dynamics of the RDS.
We supplement this well established approach with a novel, yet pivotal element, which is the inclusion of the conservation laws \cite{polettini14,polettini16,rao18:shape} of the underlying chemical network (CN) for constructing nonequilibrium thermodynamic potentials.

\emph{Theory---}The description of \cite{rao16:crnThermo} is extended to CNs endowed with a spacial structure. We consider a dilute ideal mixture of chemical species $\sigma$ that diffuse within a vessel $\caV \ni \vec r$ with impermeable walls and undergo elementary reactions $\rho$.
The abundance of some species is possibly controlled by the coupling with external chemostats (if not, the system is called closed).
Hence, the concentration $Z_{\sigma}(\vec r,t)$ of internal and chemostatted species, respectively denoted $x$ and $y$, follows the reaction-diffusion equations
\begin{align}
\partial_t Z_{\sigma} &=- \nabla \cdot \vec J_{\sigma} + {\textstyle\sum_{\rho}} \mathbb{S}^{\sigma}_{\rho} j_\rho + I_{\sigma} \, ,
	\label{rde}
\end{align}
with the Fick's diffusion current $\vec J_{\sigma}=-D_\sigma \nabla Z_\sigma$ vanishing at the boundaries of $\caV$, and the chemostatted current $ I_{\sigma}\neq 0 \; \forall y$ describing the 
rate at which the chemostatted species enter the (open) system. 

The stoichiometric matrix $ \mathbb{S}^{\sigma}_{\rho} = \nu_{-\rho}^{\sigma} - \nu_{+\rho}^{\sigma}$, i.e. the negative difference between the number of species $\sigma$ involved in the forward ($+\rho$) and backward ($-\rho$) reaction, specifies the CN topology.
Its left null vectors $\ell_\sigma^\lambda$, i.e. $\sum_\sigma \ell_\sigma^\lambda \mathbb{S}_{\rho}^{\sigma}=0$, define the components $L_{\lambda}=\sum_\sigma \ell^\lambda_\sigma Z_\sigma$, which are the global conserved quantities of the closed system: $\de_t \int_\caV \de \vec r L_{\lambda} =0$.
For this reason $\ell^\lambda_\sigma$ are called conservation laws.
Physically, they identify parts of molecules, called moieties, exchanged between species \cite{haraldsdottir16}.
When the system is opened by chemostatting, $\ell^{\lambda}_{\sigma}$ differentiate into the $\ell^{\lambda_{\rm u}}_{\sigma}$'s that are left null vectors of the submatrix of internal species $\mathbb{S}_{\rho}^{x}$, and the $\ell^{\lambda_{\rm b}}_{\sigma}$'s that are not, namely:
\begin{align}
	&{\textstyle\sum_{x}} \ell^{\lambda_{\rm u}}_{x} \mathbb{S}_{\rho}^{x}=0\,, && {\textstyle\sum_{x}} \ell^{\lambda_{\rm b}}_{x} \mathbb{S}_{\rho}^{x} \neq 0 \,.
\end{align}
Accordingly, the unbroken components $L_{\lambda_{\rm u}}=\sum_\sigma \ell^{\lambda_{\rm u}}_\sigma Z_\sigma$ remain global conserved quantities of the system, $\de_t \int_\caV \de \vec r L_{\lambda_{\rm u}} =0$, while the broken ones $L_{\lambda_{\rm b}}=\sum_\sigma \ell^{\lambda_{\rm b}}_\sigma Z_\sigma$ change over time.  In the following they will play a central role in building the nonequilibrium thermodynamics of the system. 

The net reaction current $j_\rho=j_{+\rho}-j_{-\rho}$ determines the CN dynamics.
By virtue of the mass-action kinetics assumption \cite{groot84}, each reaction current is proportional to the product of the reacting species concentrations, $j_{\pm \rho}= k_{\pm \rho} \prod_\sigma Z_\sigma^{\nu_{\pm \rho}^\sigma}$.
Thermodynamic equilibrium, characterized by homogeneous concentrations $Z_\sigma^{\rm eq}$, is reached when all external and reaction currents vanish identically, $j_\rho=I_{\sigma}=0$.
It  implies for the rate constants the \emph{local detailed balance} condition $k_{+\rho}/k_{-\rho}= \prod_\sigma ({Z^{\rm eq}_\sigma})^{\mathbb{S}^\sigma_{\rho}} $.
Such relation is taken to be valid irrespective of the system's state. The CN instead may be in a \emph{global} nonequilibrium state characterized by space-dependent concentrations $Z_\sigma (\vec r,t)$ as a result of inhomogeneous initial conditions or because of non-vanishing external currents $I_{\sigma}$. Yet, we assume it to be kept by the solvent in \emph{local} thermal equilibrium at a give temperature $T$. Therefore, the species can be assigned thermodynamic state functions, which have the known equilibrium form valid for dilute ideal mixtures, but are function of the nonequilibrium concentrations $Z_\sigma (\vec r,t)$ \cite[ch.~15]{kondepudi14}. 

A central role is played by the nonequilibrium chemical potential $\mu_{\sigma}(\vec r ) = \mu_{\sigma}^\circ + \ln Z_{\sigma}(\vec r )$ (given in units of temperature $T$ times the gas constant $R$, as any other quantity hereafter). It renders the local detailed balance in the form $k_{+\rho}/k_{-\rho}= \exp(-{\textstyle\sum_{\sigma}}   \mathbb{S}_{\rho}^{\sigma} \mu^\circ_\sigma)$, involving only the difference between the energy of formation of reactants and products. Moreover, its variation across space and between species gives the local diffusion and reaction affinity  \cite{groot84},
\begin{align}
\vec F_{\sigma}(\vec r ) &= - \nabla \mu_{\sigma}(\vec r ) \, ,
	&&f_\rho(\vec r ) = -{\textstyle\sum_{\sigma}} \mathbb{S}^\sigma_{\rho} \mu_{\sigma}(\vec r),
	\label{aff}
\end{align}
which are the thermodynamic forces driving the system.

We introduce as nonequilibrium potential the `canonical' Gibbs free energy of the system $G=\int_{\caV} \de \vec r \sum_\sigma(\mu_\sigma Z_\sigma -Z_\sigma)$   (given up to a constant). It can be expressed in terms of the equilibrium free energy $G^{\rm eq}=G(Z^{\rm eq}_\sigma)$ as
\begin{align}
G= G^{\rm eq}+ \caL(Z_\sigma \| Z^{\rm eq}_\sigma) 
\label{G}
\end{align}
introducing the \emph{relative entropy} for non-normalized concentration distributions
\begin{align}
 \caL(Z_\sigma \| Z^{\rm eq}_\sigma)=\int_\caV \de \vec r{\textstyle\sum_{\sigma}} \left[Z_\sigma \ln \frac{Z_\sigma}{Z^{\rm eq}_\sigma}- (Z_\sigma-Z^{\rm eq}_\sigma) \right].
\label{relent}
\end{align}
Akin to the  Kullback--Leibler divergence for probability densities \cite{esposito11}, \eqref{relent} quantifies the dissimilarity between two concentrations: being positive for all $Z_\sigma \neq Z^{\rm eq}_\sigma$, it implies that $G$ is always larger than its equilibrium counterpart $G^{\rm eq}$. 
Most importantly, it is minimized by the relaxation dynamics of \emph{closed} systems. This is showed evaluating the time derivative of \eqref{G} with the aid of \eqref{rde} and \eqref{aff},
\begin{align}	\label{dtG}
\dt \caL = \dt G = - \dot \Sigma_{\mathrm{dff}} -\dot  \Sigma_{\mathrm{rct}} = - \dot \Sigma \leqslant 0,
\end{align}
and recognizing the total entropy production rate (EPR) $\dot \Sigma$, split into its diffusion and reaction parts \cite{groot84}:
\begin{align}\label{EPRs}
 &\dot \Sigma_{\mathrm{dff}} = \int_\caV \de \vec r  {\textstyle\sum_{\sigma}} \mathbf{J}_{\sigma} \cdot \mathbf{F}_{\sigma}, &&\dot \Sigma_{\mathrm{rct}} = \int_\caV \de \vec r  \, {\textstyle\sum_{\rho}} j_\rho f_\rho.
 \end{align}
\begin{figure}[t]
\includegraphics[width=0.4\textwidth]{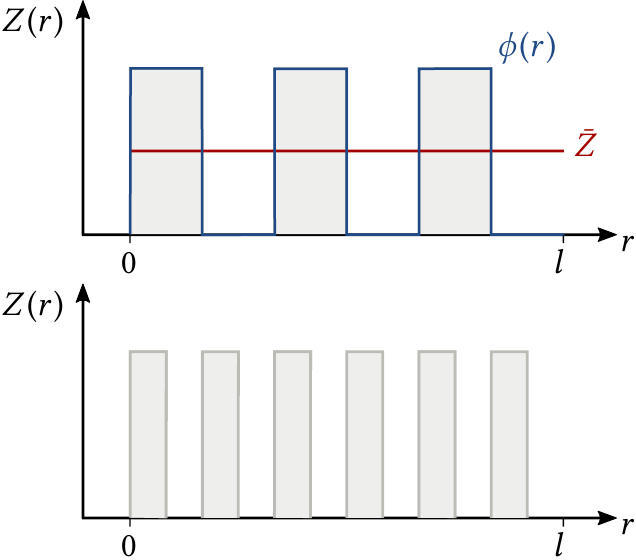}
\caption{Sketch of two patterns with equal relative entropy. Any transformation $\phi(r) \to \phi'(r)=\phi(r')$ with $|\partial r/\partial r'| =1$, corresponding to a simple rearrangement of the local concentrations, leaves $\caL(Z|Z^{\rm eq})$ unchanged. This is rooted in the lack of interactions between chemicals at the scale of RDS.}
\label{fig:pat}
\end{figure}
The relative entropy \eqref{relent} possesses some important physical features. First, in the absence of reactions it gives the total entropy produced by the diffusive expansion of concentrations. For example, consider $n_{\mathrm{A}}$ and $n_{\mathrm{B}}$ moles of inert chemicals A and B  initially placed in the volume fractions $V_{\mathrm{A}}$ and $V_{\mathrm{B}}$, respectively. They relax to homogeneous concentrations with an entropy production $-\caL= n_{\mathrm{A}} \log V_{\mathrm{A}} + n_{\mathrm{B}} \log V_{\mathrm{B}}$ that is exactly the \emph{entropy of mixing} of the two species \cite{*[][{, ch.~6.}] {ben-naim08}}. It is remarkable that diffusive dissipation and mixing entropy are thus fully described in a purely information theoretic fashion, namely as a relative entropy between concentrations.
 Second, the relative entropy between reacting concentrations $Z_\sigma(\vec r,t)= \bar Z_\sigma(t) \phi_\sigma(\vec r,t) \caV $ and arbitrary reference homogeneous concentrations  $Z_\sigma^{\rm h}$ can be split into the relative entropy between space-averaged concentrations $\bar Z_\sigma(t)= \int_\caV \de \vec r\, Z_\sigma(\vec r,t) / \caV$ and equilibrium ones $Z^{\rm eq}_\sigma$, plus the relative entropy of the normalized local modulations $\phi_\sigma(\vec r)$ around $\bar Z_\sigma$  and the flat distribution $1/\caV$:
\begin{align}
& \caL(Z_\sigma \| Z_\sigma^{\rm h})=  \caL(\bar Z_\sigma \| Z_\sigma^{\rm h}) + {\textstyle\sum_{\sigma}} \bar Z_\sigma \caL(\phi_\sigma \| 1/\caV) .
\end{align}
The positivity of relative entropy implies $\caL(Z_\sigma \|Z_\sigma^{\rm h})\geqslant  \caL(\bar Z_\sigma \| Z_\sigma^{\rm h})$, i.e. the free energy of a patterned system is always larger than its homogeneous counterpart. Third, different patterns may have the same relative entropy (see Fig.~\ref{fig:pat}) indicating that morphology and thermodynamics need not be correlated \cite{serna17}.

Notice that the conservation laws of the CN are instrumental in the derivation of \eqref{G}. 
Indeed, the equilibrium condition $ {\sum_{\sigma}} \mathbb{S}^\sigma_{\rho} \mu_{\sigma}^{\rm eq}=0$ corresponding to null reaction affinities, $f_\rho=0$, implies that $ \mu_{\sigma}^{\rm eq}$ is a linear combination of the conservation laws $\ell^{\lambda}_\sigma$. This entails $ \int_{\caV} \de \vec r\sum_\sigma \mu_\sigma^{\rm eq} \partial_t Z_\sigma =0$, which yields in turn the decomposition \eqref{G} when time integrating along a relaxation dynamics leading from $Z_\sigma$ to $Z_\sigma^{\rm eq}$. 

Moreover,  the conservation laws are the passkey to construct the correct nonequilibrium thermodynamic potential for \emph{open} systems. For the latter, 
an additional term appears when taking the time derivative of $G$ due to the external current in \eqref{rde}, 
\begin{equation}
	\dot W_{\mathrm{chem}} = \int_\caV \de \vec r  \, {\textstyle \sum_{\sigma}} \mu_{\sigma}(\vec r ) I_{\sigma}(\vec r ) \, ,
	\label{Wc}
\end{equation}
which defines the chemical work performed by the chemostats. The second law \eqref{dtG} thus attains the new form 
   \vspace{-0.1cm}
\begin{align}
 \dot{W}_{\mathrm{chem}} -\dt G =\dot \Sigma  \geqslant 0,
 \end{align}
 where the EPR $\dot \Sigma$ is still given by the two contributions of Eq.~\eqref{EPRs}.
 Consequently, $G$ is no longer minimized due to the break of conservations laws. Similarly to equilibrium thermodynamics when passing from canonical to grand canonical ensembles, one needs to transform the free energy  $G$ subtracting the energetic contributions of matter exchanged with the reservoirs \cite{alberty03}. This amounts to the moieties of the broken components $M_{\yp} = {\textstyle\sum_{\lambda_{\mathrm{b}}}} {\ell_{\yp}^{\lambda_{\mathrm{b}}}}^{-1} \int \de \vec r \, L_{\lambda_{\mathrm{b}}}(\vec r)$ entering those chemostats $\yp$ that break all conservation laws, times the reference values of their chemical potential $\muRef $ (which simplifies to $\mu_{\yp}$ for homogeneous chemostats). The so obtained semigrand Gibbs free energy,
\begin{align}
\caG= G- {\textstyle \sum_{\yp}} \muRef   M_{\yp} \, ,
\end{align} 
encodes CN-specific topological and spatial features thanks to the freedom in the choice of $\yp$ and $\muRef $. 
This allows one to split the EPR
\begin{equation}
	\dot{W}_{\mathrm{driv}}
	+  \dot W_{\mathrm{nc}} 
	- \dt \caG
	= 	 \dot\Sigma \, ,
	\label{EPsplit}
\end{equation}
in terms of the driving and the nonconservative chemical work rate, respectively,
\begin{align}
	\hspace{-1em}
	\dot{W}_{\mathrm{driv}} & = - {\textstyle\sum_{\yp}} \dt \muRef M_{\yp} \, , &
	\dot W_{\mathrm{nc}} & = {\textstyle \sum_{y}} \int_{\caV} \de \vec r I_{y} \mathcal{F}_{y} \, .
\end{align}
The former results from time-dependent manipulations of the reference chemostats $\yp$, while the latter gives the cost of sustaining chemical flows by means of the forces $\mathcal{F}_{y}(\vec r) =
\mu_{y}(\vec r) - {\textstyle\sum_{\yp}} \muRef {\textstyle\sum_{\lambda_{\mathrm{b}}}} {\ell_{\yp}^{\lambda_{\mathrm{b}}}}^{-1}\ell_{y}^{\lambda_{\mathrm{b}}}$ measured with respect to the reference chemical potentials $\muRef$.
Eq.~\eqref{EPsplit} is a major result of this Letter and can be verified by direct substitution. 

In absence of driving ($\dt \muRef$=0) and nonconservative forcing ($\caF_y=0$) it simplifies to $\dt \caG=- \dot\Sigma\leqslant0 $,
which proves that the CN, despite being \emph{open}, relaxes to equilibrium by minimizing the free energy $\caG$. Moreover, for a generic open CN, the decomposition of $\caG$ corresponding to \eqref{G}, i.e. $\caG- \caG^{\mathrm{eq}} = \caL (Z_x \| Z^{\rm eq}_x)\geqslant0 $, and a time integral between two nonequilibrium states connected by an arbitrary manipulation turn \eqref{EPsplit} into a \emph{nonequilibrium Landauer principle} \cite{esposito11} for RDS, 
 \begin{align}
W_{\mathrm{driv}}
+  W_{\mathrm{nc}} 
	- \Delta \caG_{\mathrm{eq}}
	 \geqslant \Delta \caL(Z_x \| Z^{\rm eq}_x)  \, .
	\label{Land}
 \end{align}
The latter states that the dissipative work spent to manipulate the CN is bounded by the variation in relative entropy between the boundary states and their respective equilibria attained by stopping the driving and zeroing the forcing.

\emph{Turing pattern in the Brusselator model---}As first proposed by A.~Turing in his seminal paper \cite{turing90}, RDS undergo a spatial symmetry braking leading to a stationary pattern when at least two chemical species react nonlinearly and their diffusivities differ substantially.
 A minimal system that captures these essential features is the Brusselator model \cite{prigogine68} in one spatial dimension. Here the concentrations of two chemical species, an activator $Z_{X_1}=x_1(r,t)$ and an inhibitor  $Z_{X_2}=x_2(r,t)$, evolve in time and space $r \in [0,l]$ according to the RDS \eqref{rde} for the network depicted in Fig.~\ref{fig:G}, namely,
\begin{widetext}
\begin{align}
\label{brus}
\partial_t 
\begin{pmatrix}
x_1\\
x_2
\end{pmatrix}
=
\underbrace{\begin{pmatrix}
 k_1 y_1 - k_{-1}x_1 - k_2 y_2 x_1 + k_{-2} y_3 x_2 + k_{3} x_1^2 x_2 - k_{-3} x_1^3 -k_{4} x_1 + k_{-4} y_4 + D_{x_1} \partial_r^2 x_1, \\
  k_2 y_2 x_1 - k_{-2} y_3 x_2 - k_{3} x_1^2 x_2 + k_{-3} x_1^3 + D_{x_2} \partial_r^2 x_2
\end{pmatrix}}_{\text{\small  $ = \caJ(x_1,x_2)$}},
\end{align}
\end{widetext}
The $y_1, y_2, y_3$ and $y_4$ are the homogeneous concentrations of the chemostatted species and the diffusivities satisfy the Turing condition $D_{x_1} \ll D_{x_2}$. 
Equation \eqref{brus} admits a homogeneous stationary solution $(x_1^\mathrm{h},x_2^\mathrm{h})^\mathsf{T}$ 
that becomes unstable for $y_2 \geqslant y_2^\mathrm{c}$, so that a sinusoidal pattern with wavelength $q_\mathrm{c}$ and amplitude proportional to the (in general complex) function  $A(r,t)$ starts developing around the space-averaged concentrations $\bar x(t)$ \cite{kondepudi14}:
\begin{equation}\label{sol}
	\hspace{-1em}
\begin{pmatrix}
x_1(r,t)\\
x_2(r,t)
\end{pmatrix}
\!=\! 
\begin{pmatrix}
\bar x_1(t) \\
\bar x_2(t)
\end{pmatrix}
+ 
\begin{pmatrix}
1 \\
u_{x_2}
\end{pmatrix}
 \left(A(r,t) e^{iq_\mathrm{c} r} + c.c. \right).
\end{equation}
\begin{figure}[t]
\begin{tikzpicture}
\begin{axis}[name=mainplot, 
xmin=ln(2.55),xmax=ln(2.75),ymin=-31.65,ymax=-31.45,
xlabel=$\mu_{Y_2}-\mu_{Y_2}^\circ$, ylabel=$\caG$, ylabel style={rotate=-90}] 
\addplot[color=scarlet2,dotted ,style={ultra thick} ] coordinates{(
(ln 2.55, -31.4254)(ln 2.552, -31.4278)(ln 2.554, -31.4303)(ln 2.556, 
-31.4327)(ln 2.558, -31.4352)(ln 2.56, -31.4376)(ln 2.562, -31.4401) 
(ln 2.564, -31.4425)(ln 2.566, -31.4449)(ln 2.568, -31.4474)(ln 2.57, 
-31.4498)(ln 2.572, -31.4523)(ln 2.574, -31.4547)(ln 2.576, -31.4572) 
(ln 2.578, -31.4596)(ln 2.58, -31.462)(ln 2.582, -31.4645)(ln 2.584, 
-31.4669)(ln 2.586, -31.4693)(ln 2.588, -31.4718)(ln 2.59, -31.4742) 
(ln 2.592, -31.4767)(ln 2.594, -31.4791)(ln 2.596, -31.4815)(ln 2.598, 
-31.4839)(ln 2.6, -31.4864)(ln 2.602, -31.4888)(ln 2.604, -31.4912) 
(ln 2.606, -31.4937)(ln 2.608, -31.4961)(ln 2.61, -31.4985)(ln 2.612, 
-31.501)(ln 2.614, -31.5034)(ln 2.616, -31.5058)(ln 2.618, -31.5082) 
(ln 2.62, -31.5106)(ln 2.622, -31.5131)(ln 2.624, -31.5155)(ln 2.626, 
-31.5179)(ln 2.628, -31.5203)(ln 2.63, -31.5228)(ln 2.632, -31.5252) 
(ln 2.634, -31.5276)(ln 2.636, -31.53)(ln 2.638, -31.5324)(ln 2.64, 
-31.5348)(ln 2.642, -31.5373)(ln 2.644, -31.5397)(ln 2.646, -31.5421) 
(ln 2.648, -31.5445)(ln 2.65, -31.5469)(ln 2.652, -31.5493)(ln 2.654, 
-31.5517)(ln 2.656, -31.5541)(ln 2.658, -31.5566)(ln 2.66, -31.559) 
(ln 2.662, -31.5614)(ln 2.664, -31.5638)(ln 2.666, -31.5662)(ln 2.668, 
-31.5686)(ln 2.67, -31.571)(ln 2.672, -31.5734)(ln 2.674, -31.5758) 
(ln 2.676, -31.5782)(ln 2.678, -31.5806)(ln 2.68, -31.583)(ln 2.682, 
-31.5854)(ln 2.684, -31.5878)(ln 2.686, -31.5902)(ln 2.688, -31.5926) 
(ln 2.69, -31.595)(ln 2.692, -31.5974)(ln 2.694, -31.5998)(ln 2.696, 
-31.6022)(ln 2.698, -31.6046)(ln 2.7, -31.607)(ln 2.702, -31.6094) 
(ln 2.704, -31.6118)(ln 2.706, -31.6141)(ln 2.708, -31.6165)(ln 2.71, 
-31.6189)(ln 2.712, -31.6213)(ln 2.714, -31.6237)(ln 2.716, -31.6261) 
(ln 2.718, -31.6285)(ln 2.72, -31.6309)(ln 2.722, -31.6332)(ln 2.724, 
-31.6356)(ln 2.726, -31.638)(ln 2.728, -31.6404)(ln 2.73, -31.6428) 
(ln 2.732, -31.6452)(ln 2.734, -31.6475)(ln 2.736, -31.6499)(ln 2.738, 
-31.6523)(ln 2.74, -31.6547)(ln 2.742, -31.6571)(ln 2.744, -31.6594) 
(ln 2.746, -31.6618)(ln 2.748, -31.6642)(ln 2.75, -31.6666)
};
\addplot[color=skyblue2,style={ultra thick}] coordinates{
(ln 2.55, -31.4254) (ln 2.552, -31.4278) (ln 2.554, -31.4303) (ln 2.556, -31.4327) (ln 2.558, -31.4351) (ln 2.56, -31.4376) (ln 2.562, -31.44)  (ln 2.564, -31.4425) (ln 2.566, -31.4449) (ln 2.568, -31.4474) (ln 2.57,  
-31.4498) (ln 2.572, -31.4523) (ln 2.574, -31.4547) (ln 2.576, -31.4571)  
(ln 2.578, -31.4596) (ln 2.58, -31.462) (ln 2.582, -31.4645) (ln 2.584,  
-31.4669) (ln 2.586, -31.4693) (ln 2.588, -31.4718) (ln 2.59, -31.4742)  
(ln 2.592, -31.4766) (ln 2.594, -31.4791) (ln 2.596, -31.4815) (ln 2.598,  
-31.4839) (ln 2.6, -31.4864) (ln 2.602, -31.4888) (ln 2.604, -31.4912)  
(ln 2.606, -31.4937) (ln 2.608, -31.4961) (ln 2.61, -31.4985) (ln 2.612,  
-31.5009) (ln 2.614, -31.5034) (ln 2.616, -31.5058) (ln 2.618, -31.5082)  
(ln 2.62, -31.5106) (ln 2.622, -31.5131) (ln 2.624, -31.5155) (ln 2.626,  
-31.5179) (ln 2.628, -31.5203) (ln 2.63, -31.5227) (ln 2.632, -31.5252)  
(ln 2.634, -31.5276) (ln 2.636, -31.53) (ln 2.638, -31.5324) (ln 2.64,  
-31.5348) (ln 2.642, -31.5372) (ln 2.644, -31.5397) (ln 2.646, -31.5421)  
(ln 2.648, -31.5445) (ln 2.65, -31.5469) (ln 2.652, -31.5493) (ln 2.654,  
-31.5517) (ln 2.656, -31.5541) (ln 2.658, -31.5565) (ln 2.66, -31.5589)  
(ln 2.662, -31.5614) (ln 2.664, -31.5638) (ln 2.666, -31.5656) (ln 2.668,  
-31.5669) (ln 2.67, -31.5683) (ln 2.672, -31.5696) (ln 2.674, -31.571)  
(ln 2.676, -31.5723) (ln 2.678, -31.5736) (ln 2.68, -31.575) (ln 2.682,  
-31.5763) (ln 2.684, -31.5776) (ln 2.686, -31.579) (ln 2.688, -31.5803)  
(ln 2.69, -31.5816) (ln 2.692, -31.5829) (ln 2.694, -31.5843) (ln 2.696,  
-31.5856) (ln 2.698, -31.5869) (ln 2.7, -31.5882) (ln 2.702, -31.5895)  
(ln 2.704, -31.5909) (ln 2.706, -31.5922) (ln 2.708, -31.5935) (ln 2.71,  
-31.5948) (ln 2.712, -31.5961) (ln 2.714, -31.5974) (ln 2.716, -31.5988)  
(ln 2.718, -31.6001) (ln 2.72, -31.6014) (ln 2.722, -31.6027) (ln 2.724,  
-31.604) (ln 2.726, -31.6053) (ln 2.728, -31.6066) (ln 2.73, -31.6079)  
(ln 2.732, -31.6092) (ln 2.734, -31.6105) (ln 2.736, -31.6118) (ln 2.738,  
-31.6131) (ln 2.74, -31.6144) (ln 2.742, -31.6157) (ln 2.744, -31.617)  
(ln 2.746, -31.6183) (ln 2.748, -31.6196) (ln 2.75, -31.6209)
};
\addplot[color=black, mark=star, mark size=2.8pt, only marks, style=thick] coordinates{
(ln 2.6649467391417865, -31.56491295236406)
(ln 2.6849467391417865, -31.578406177)
(ln 2.7049467391417865, -31.591734697905533)
(ln 2.7249467391417865, -31.60528716318163)
(ln 2.7449467391417866, -31.61862450977896)
(ln 2.7649467391417866, -31.632083808250275)
(ln 2.7849467391417866, -31.645712344868016)
(ln 2.8049467391417866, -31.659030774918335)
(ln 2.8249467391417866, -31.672518125070848)
(ln 2.8449467391417866, -31.686014656696152)
(ln 2.8649467391417867, -31.699520408670825)
(ln 2.8849467391417867, -31.713260424502128)
(ln 2.9049467391417867, -31.726797568521402)
(ln 2.9249467391417863, -31.74009391441306)
(ln 2.398452065227608, -31.23741069679327)
(ln 2.418452065227608, -31.262483691279396)
(ln 2.438452065227608, -31.287473991072112)
(ln 2.458452065227608, -31.31238225634381)
(ln 2.478452065227608, -31.337209191114972)
(ln 2.498452065227608, -31.361955425148786)
(ln 2.518452065227608, -31.386621606938338)
(ln 2.538452065227608, -31.411208367893593)
(ln 2.558452065227608, -31.435716342969464)
(ln 2.578452065227608, -31.46014614819357)
(ln 2.598452065227608, -31.48449838302214)
(ln 2.618452065227608, -31.508773641284765)
(ln 2.6384520652276082, -31.5329725226587)
(ln 2.658452065227608, -31.557095600552525)};
 \end{axis}
\begin{axis}[anchor=south west, xshift=0.3cm, yshift=0.25cm, xmin=ln(2.55), xmax=ln(2.75),width=0.23\textwidth, ylabel style={rotate=-90}, ylabel=$\partial \caG/\partial \mu_{Y_2}$,style={font=\scriptsize}, ylabel near ticks, yticklabel pos=right, xlabel near ticks, xticklabel pos=upper]
\addplot[color=scarlet2,dotted,style={ultra thick}] coordinates {(ln 2.55, -3.12399) (ln 2.551, -3.12471) (ln 2.552, -3.12544) (ln 2.553, 
-3.12616) (ln 2.554, -3.12689) (ln 2.555, -3.12761) (ln 2.556, -3.12834) 
(ln 2.557, -3.12906) (ln 2.558, -3.12978) (ln 2.559, -3.13051) (ln 2.56, 
-3.13123) (ln 2.561, -3.13195) (ln 2.562, -3.13268) (ln 2.563, -3.1334) 
(ln 2.564, -3.13412) (ln 2.565, -3.13484) (ln 2.566, -3.13557) (ln 2.567, 
-3.13629) (ln 2.568, -3.13701) (ln 2.569, -3.13773) (ln 2.57, -3.13845) 
(ln 2.571, -3.13917) (ln 2.572, -3.13989) (ln 2.573, -3.14061) (ln 2.574, 
-3.14134) (ln 2.575, -3.14206) (ln 2.576, -3.14278) (ln 2.577, -3.1435) 
(ln 2.578, -3.14422) (ln 2.579, -3.14493) (ln 2.58, -3.14565) (ln 2.581, 
-3.14637) (ln 2.582, -3.14709) (ln 2.583, -3.14781) (ln 2.584, -3.14853) 
(ln 2.585, -3.14925) (ln 2.586, -3.14997) (ln 2.587, -3.15068) (ln 2.588, 
-3.1514) (ln 2.589, -3.15212) (ln 2.59, -3.15284) (ln 2.591, -3.15355) 
(ln 2.592, -3.15427) (ln 2.593, -3.15499) (ln 2.594, -3.1557) (ln 2.595, 
-3.15642) (ln 2.596, -3.15714) (ln 2.597, -3.15785) (ln 2.598, -3.15857) 
(ln 2.599, -3.15928) (ln 2.6, -3.16) (ln 2.601, -3.16072) (ln 2.602, 
-3.16143) (ln 2.603, -3.16215) (ln 2.604, -3.16286) (ln 2.605, -3.16357) 
(ln 2.606, -3.16429) (ln 2.607, -3.165) (ln 2.608, -3.16572) (ln 2.609, 
-3.16643) (ln 2.61, -3.16714) (ln 2.611, -3.16786) (ln 2.612, -3.16857) 
(ln 2.613, -3.16928) (ln 2.614, -3.17) (ln 2.615, -3.17071) (ln 2.616, 
-3.17142) (ln 2.617, -3.17213) (ln 2.618, -3.17285) (ln 2.619, -3.17356) 
(ln 2.62, -3.17427) (ln 2.621, -3.17498) (ln 2.622, -3.17569) (ln 2.623, 
-3.1764) (ln 2.624, -3.17711) (ln 2.625, -3.17783) (ln 2.626, -3.17854) 
(ln 2.627, -3.17925) (ln 2.628, -3.17996) (ln 2.629, -3.18067) (ln 2.63, 
-3.18138) (ln 2.631, -3.18209) (ln 2.632, -3.18279) (ln 2.633, -3.1835) 
(ln 2.634, -3.18421) (ln 2.635, -3.18492) (ln 2.636, -3.18563) (ln 2.637, 
-3.18634) (ln 2.638, -3.18705) (ln 2.639, -3.18775) (ln 2.64, -3.18846) 
(ln 2.641, -3.18917) (ln 2.642, -3.18988) (ln 2.643, -3.19059) (ln 2.644, 
-3.19129) (ln 2.645, -3.192) (ln 2.646, -3.19271) (ln 2.647, -3.19341) 
(ln 2.648, -3.19412) (ln 2.649, -3.19482) (ln 2.65, -3.19553) (ln 2.651, 
-3.19624) (ln 2.652, -3.19694) (ln 2.653, -3.19765) (ln 2.654, -3.19835) 
(ln 2.655, -3.19906) (ln 2.656, -3.19976) (ln 2.657, -3.20047) (ln 2.658, 
-3.20117) (ln 2.659, -3.20188) (ln 2.66, -3.20258) (ln 2.661, -3.20328) 
(ln 2.662, -3.20399) (ln 2.663, -3.20469) (ln 2.664, -3.20539) (ln 2.665, 
-3.2061) (ln 2.666, -3.2068) (ln 2.667, -3.2075) (ln 2.668, -3.20821) 
(ln 2.669, -3.20891) (ln 2.67, -3.20961) (ln 2.671, -3.21031) (ln 2.672, 
-3.21101) (ln 2.673, -3.21172) (ln 2.674, -3.21242) (ln 2.675, -3.21312) 
(ln 2.676, -3.21382) (ln 2.677, -3.21452) (ln 2.678, -3.21522) (ln 2.679, 
-3.21592) (ln 2.68, -3.21662) (ln 2.681, -3.21732) (ln 2.682, -3.21802) 
(ln 2.683, -3.21872) (ln 2.684, -3.21942) (ln 2.685, -3.22012) (ln 2.686, 
-3.22082) (ln 2.687, -3.22152) (ln 2.688, -3.22222) (ln 2.689, -3.22292) 
(ln 2.69, -3.22361) (ln 2.691, -3.22431) (ln 2.692, -3.22501) (ln 2.693, 
-3.22571) (ln 2.694, -3.22641) (ln 2.695, -3.2271) (ln 2.696, -3.2278) 
(ln 2.697, -3.2285) (ln 2.698, -3.2292) (ln 2.699, -3.22989) (ln 2.7, 
-3.23059) (ln 2.701, -3.23129) (ln 2.702, -3.23198) (ln 2.703, -3.23268) 
(ln 2.704, -3.23337) (ln 2.705, -3.23407) (ln 2.706, -3.23477) (ln 2.707, 
-3.23546) (ln 2.708, -3.23616) (ln 2.709, -3.23685) (ln 2.71, -3.23755) 
(ln 2.711, -3.23824) (ln 2.712, -3.23893) (ln 2.713, -3.23963) (ln 2.714, 
-3.24032) (ln 2.715, -3.24102) (ln 2.716, -3.24171) (ln 2.717, -3.2424) 
(ln 2.718, -3.2431) (ln 2.719, -3.24379) (ln 2.72, -3.24448) (ln 2.721, 
-3.24518) (ln 2.722, -3.24587) (ln 2.723, -3.24656) (ln 2.724, -3.24725) 
(ln 2.725, -3.24794) (ln 2.726, -3.24864) (ln 2.727, -3.24933) (ln 2.728, 
-3.25002) (ln 2.729, -3.25071) (ln 2.73, -3.2514) (ln 2.731, -3.25209) 
(ln 2.732, -3.25278) (ln 2.733, -3.25347) (ln 2.734, -3.25416) (ln 2.735, 
-3.25485) (ln 2.736, -3.25554) (ln 2.737, -3.25623) (ln 2.738, -3.25692) 
(ln 2.739, -3.25761) (ln 2.74, -3.2583) (ln 2.741, -3.25899) (ln 2.742, 
-3.25968) (ln 2.743, -3.26037) (ln 2.744, -3.26106) (ln 2.745, -3.26175) 
(ln 2.746, -3.26243) (ln 2.747, -3.26312) (ln 2.748, -3.26381) (ln 2.749, 
-3.2645) (ln 2.75, -3.26518)};
\addplot[color=skyblue2,style={ultra thick}] coordinates {
(ln 2.55, -3.12402) (ln 2.551, -3.12474) (ln 2.552, -3.12547) (ln 2.553, 
-3.12619) (ln 2.554, -3.12692) (ln 2.555, -3.12764) (ln 2.556, -3.12836) 
(ln 2.557, -3.12909) (ln 2.558, -3.12981) (ln 2.559, -3.13053) (ln 2.56, 
-3.13126) (ln 2.561, -3.13198) (ln 2.562, -3.1327) (ln 2.563, -3.13343) 
(ln 2.564, -3.13415) (ln 2.565, -3.13487) (ln 2.566, -3.13559) (ln 2.567, 
-3.13631) (ln 2.568, -3.13704) (ln 2.569, -3.13776) (ln 2.57, -3.13848) 
(ln 2.571, -3.1392) (ln 2.572, -3.13992) (ln 2.573, -3.14064) (ln 2.574, 
-3.14136) (ln 2.575, -3.14208) (ln 2.576, -3.1428) (ln 2.577, -3.14352) 
(ln 2.578, -3.14424) (ln 2.579, -3.14496) (ln 2.58, -3.14568) (ln 2.581, 
-3.1464) (ln 2.582, -3.14712) (ln 2.583, -3.14784) (ln 2.584, -3.14856) 
(ln 2.585, -3.14928) (ln 2.586, -3.14999) (ln 2.587, -3.15071) (ln 2.588, 
-3.15143) (ln 2.589, -3.15215) (ln 2.59, -3.15286) (ln 2.591, -3.15358) 
(ln 2.592, -3.1543) (ln 2.593, -3.15502) (ln 2.594, -3.15573) (ln 2.595, 
-3.15645) (ln 2.596, -3.15716) (ln 2.597, -3.15788) (ln 2.598, -3.1586) 
(ln 2.599, -3.15931) (ln 2.6, -3.16003) (ln 2.601, -3.16074) (ln 2.602, 
-3.16146) (ln 2.603, -3.16217) (ln 2.604, -3.16289) (ln 2.605, -3.1636) 
(ln 2.606, -3.16432) (ln 2.607, -3.16503) (ln 2.608, -3.16574) (ln 2.609, 
-3.16646) (ln 2.61, -3.16717) (ln 2.611, -3.16789) (ln 2.612, -3.1686) 
(ln 2.613, -3.16931) (ln 2.614, -3.17002) (ln 2.615, -3.17074) (ln 2.616, 
-3.17145) (ln 2.617, -3.17216) (ln 2.618, -3.17287) (ln 2.619, -3.17359) 
(ln 2.62, -3.1743) (ln 2.621, -3.17501) (ln 2.622, -3.17572) (ln 2.623, 
-3.17643) (ln 2.624, -3.17714) (ln 2.625, -3.17785) (ln 2.626, -3.17856) 
(ln 2.627, -3.17927) (ln 2.628, -3.17998) (ln 2.629, -3.18069) (ln 2.63, 
-3.1814) (ln 2.631, -3.18211) (ln 2.632, -3.18282) (ln 2.633, -3.18353) 
(ln 2.634, -3.18424) (ln 2.635, -3.18495) (ln 2.636, -3.18566) (ln 2.637, 
-3.18637) (ln 2.638, -3.18707) (ln 2.639, -3.18778) (ln 2.64, -3.18849) 
(ln 2.641, -3.1892) (ln 2.642, -3.18991) (ln 2.643, -3.19061) (ln 2.644, 
-3.19132) (ln 2.645, -3.19203) (ln 2.646, -3.19273) (ln 2.647, -3.19344) 
(ln 2.648, -3.19415) (ln 2.649, -3.19485) (ln 2.65, -3.19556) (ln 2.651, 
-3.19626) (ln 2.652, -3.19697) (ln 2.653, -3.19768) (ln 2.654, -3.19838) 
(ln 2.655, -3.19909) (ln 2.656, -3.19979) (ln 2.657, -3.20049) (ln 2.658, 
-3.2012) (ln 2.659, -3.2019) (ln 2.66, -3.20261) (ln 2.661, -3.20331) 
(ln 2.662, -3.20402) (ln 2.663, -3.20472) (ln 2.664, -3.20542) (ln 2.665, 
-1.78773) (ln 2.666, -1.78759) (ln 2.667, -1.78744) (ln 2.668, -1.7873) 
(ln 2.669, -1.78716) (ln 2.67, -1.78702) (ln 2.671, -1.78688) (ln 2.672, 
-1.78674) (ln 2.673, -1.7866) (ln 2.674, -1.78646) (ln 2.675, -1.78632) 
(ln 2.676, -1.78618) (ln 2.677, -1.78604) (ln 2.678, -1.7859) (ln 2.679, 
-1.78576) (ln 2.68, -1.78562) (ln 2.681, -1.78548) (ln 2.682, -1.78535) 
(ln 2.683, -1.78521) (ln 2.684, -1.78507) (ln 2.685, -1.78493) (ln 2.686, 
-1.7848) (ln 2.687, -1.78466) (ln 2.688, -1.78452) (ln 2.689, -1.78439) 
(ln 2.69, -1.78425) (ln 2.691, -1.78411) (ln 2.692, -1.78398) (ln 2.693, 
-1.78384) (ln 2.694, -1.78371) (ln 2.695, -1.78357) (ln 2.696, -1.78344) 
(ln 2.697, -1.7833) (ln 2.698, -1.78317) (ln 2.699, -1.78304) (ln 2.7, 
-1.7829) (ln 2.701, -1.78277) (ln 2.702, -1.78263) (ln 2.703, -1.7825) 
(ln 2.704, -1.78237) (ln 2.705, -1.78224) (ln 2.706, -1.7821) (ln 2.707, 
-1.78197) (ln 2.708, -1.78184) (ln 2.709, -1.78171) (ln 2.71, -1.78157) 
(ln 2.711, -1.78144) (ln 2.712, -1.78131) (ln 2.713, -1.78118) (ln 2.714, 
-1.78105) (ln 2.715, -1.78092) (ln 2.716, -1.78079) (ln 2.717, -1.78066) 
(ln 2.718, -1.78053) (ln 2.719, -1.7804) (ln 2.72, -1.78027) (ln 2.721, 
-1.78014) (ln 2.722, -1.78001) (ln 2.723, -1.77988) (ln 2.724, -1.77975) 
(ln 2.725, -1.77962) (ln 2.726, -1.77949) (ln 2.727, -1.77936) (ln 2.728, 
-1.77923) (ln 2.729, -1.7791) (ln 2.73, -1.77898) (ln 2.731, -1.77885) 
(ln 2.732, -1.77872) (ln 2.733, -1.77859) (ln 2.734, -1.77846) (ln 2.735, 
-1.77834) (ln 2.736, -1.77821) (ln 2.737, -1.77808) (ln 2.738, -1.77795) 
(ln 2.739, -1.77783) (ln 2.74, -1.7777) (ln 2.741, -1.77757) (ln 2.742, 
-1.77745) (ln 2.743, -1.77732) (ln 2.744, -1.77719) (ln 2.745, -1.77707) 
(ln 2.746, -1.77694) (ln 2.747, -1.77682) (ln 2.748, -1.77669) (ln 2.749, 
-1.77656) (ln 2.75, -1.77644)};
\end{axis}
\node[inner sep=0pt] (CRN) at (5.3,4.3)
{\includegraphics[width=.15\textwidth]{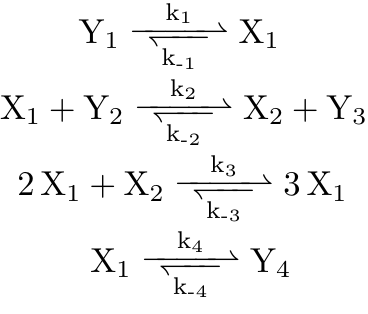}};
\end{tikzpicture}
\caption{Nonequilibrium semigrand Gibbs free energy $\caG$  for the Brusselator model as a function of the chemical potential of the chemostatted species $Z_{Y_2}$, obtained by the analytic stationary solution of the amplitude equation.
	To define $\caG$ we choose $y_1$ and $y_2$ as the reference chemostats breaking the two components $L_1=x_1+x_2+y_1+y_4$ and $L_2=y_2+y_3$.
	The dotted line represents the free energy $\caG$ in the unstable homogeneous system before the pattern growth.
	Symbols  (\textcolor{black}{$\star$}) result from numerical integration of \eqref{brus}.
	\emph{Inset}: the derivative $\partial \caG/\partial \mu_{Y_2}$ displays a discontinuity at $y_2^{\mathrm{c}}\simeq 2.66$.}
	\label{fig:G}
\end{figure}
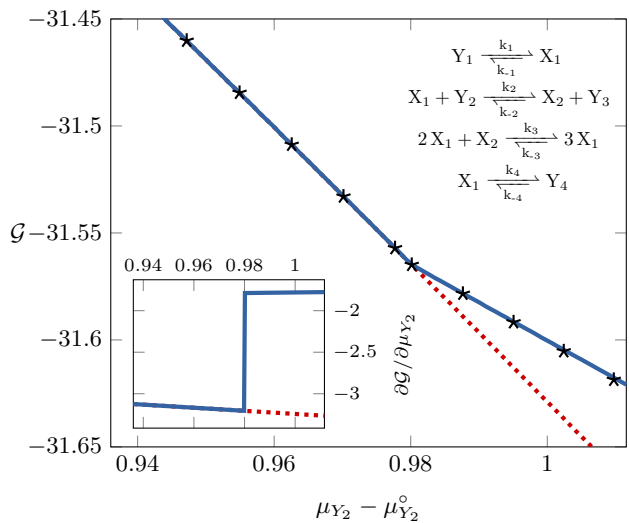
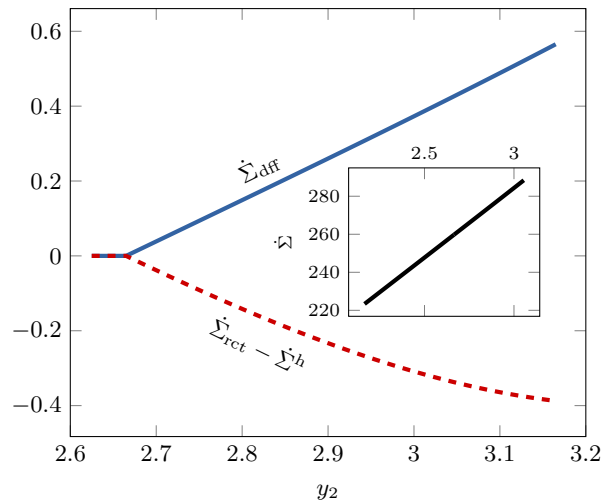
\begin{figure}
\begin{tikzpicture}
\begin{axis}[xmin=2.6,xmax=3.2,
xlabel=$y_2$] 
\addplot[color=skyblue2, no markers,style={ultra thick}] coordinates{
(2.62495, 0)(2.63495, 0)(2.64495, 0)(2.65495, 0)(2.66495, 0.)(2.67495, 0.0110184)(2.68495, 0.0220347)(2.69495, 
  0.0330498)(2.70495, 0.0440646)(2.71495, 0.05508)(2.72495, 
  0.0660967)(2.73495, 0.0771157)(2.74495, 0.0881377)(2.75495, 
  0.0991637)(2.76495, 0.110194)(2.77495, 0.121231)(2.78495, 
  0.132273)(2.79495, 0.143323)(2.80495, 0.154381)(2.81495, 
  0.165448)(2.82495, 0.176525)(2.83495, 0.187612)(2.84495, 
  0.198711)(2.85495, 0.209822)(2.86495, 0.220946)(2.87495, 
  0.232084)(2.88495, 0.243237)(2.89495, 0.254405)(2.90495, 
  0.26559)(2.91495, 0.276792)(2.92495, 0.288013)(2.93495, 
  0.299253)(2.94495, 0.310513)(2.95495, 0.321793)(2.96495, 
  0.333096)(2.97495, 0.344421)(2.98495, 0.355771)(2.99495, 
  0.367145)(3.00495, 0.378544)(3.01495, 0.389971)(3.02495, 
  0.401425)(3.03495, 0.412908)(3.04495, 0.42442)(3.05495, 
  0.435964)(3.06495, 0.447539)(3.07495, 0.459148)(3.08495, 
  0.470791)(3.09495, 0.482469)(3.10495, 0.494184)(3.11495, 
  0.505937)(3.12495, 0.517729)(3.13495, 0.529561)(3.14495, 
  0.541435)(3.15495, 0.553352)(3.16495, 0.565314)
  };
  \addplot[color=scarlet2, dashed, no markers,style={ultra thick}] coordinates{
(2.62495, 0)(2.63495, 0)(2.64495, 0)(2.65495, 0)(2.66495, 0)(2.67495, -0.0109957)(2.68495, -0.0219266) 
(2.69495, -0.0327882)(2.70495, -0.043576)(2.71495, -0.0542856) 
(2.72495, -0.0649127)(2.73495, -0.075453)(2.74495, -0.0859022) 
(2.75495, -0.0962559)(2.76495, -0.10651)(2.77495, -0.11666) 
(2.78495, -0.126702)(2.79495, -0.136632)(2.80495, -0.146446) 
(2.81495, -0.156139)(2.82495, -0.165708)(2.83495, -0.175147) 
(2.84495, -0.184454)(2.85495, -0.193624)(2.86495, -0.202653) 
(2.87495, -0.211536)(2.88495, -0.220271)(2.89495, -0.228852) 
(2.90495, -0.237276)(2.91495, -0.245539)(2.92495, -0.253636) 
(2.93495, -0.261564)(2.94495, -0.269318)(2.95495, -0.276894) 
(2.96495, -0.284288)(2.97495, -0.291496)(2.98495, -0.298515) 
(2.99495, -0.305338)(3.00495, -0.311963)(3.01495, -0.318386) 
(3.02495, -0.324601)(3.03495, -0.330605)(3.04495, -0.336393) 
(3.05495, -0.341961)(3.06495, -0.347305)(3.07495, -0.35242) 
(3.08495, -0.357301)(3.09495, -0.361944)(3.10495, -0.366345) 
(3.11495, -0.370499)(3.12495, -0.3744)(3.13495, -0.378045) 
(3.14495, -0.381428)(3.15495, -0.384544)(3.16495, -0.387388)
  };
\end{axis};
\begin{axis}[anchor=south west, xshift=3.7cm, yshift=1.6cm,width=0.23\textwidth, ylabel=$\dot \Sigma$, ylabel style={rotate=90}, style={font=\scriptsize}, ylabel near ticks, yticklabel pos=left, xlabel near ticks, xticklabel pos=upper]
\addplot[color=black, no markers,style={ultra thick}] coordinates{
(2.16495, 223.354) (2.17495, 224.08) (2.18495, 224.806) (2.19495, 
  225.533) (2.20495, 226.259) (2.21495, 226.986) (2.22495, 
  227.713) (2.23495, 228.44) (2.24495, 229.167) (2.25495, 
  229.893) (2.26495, 230.621) (2.27495, 231.348) (2.28495, 
  232.075) (2.29495, 232.802) (2.30495, 233.53) (2.31495, 
  234.257) (2.32495, 234.985) (2.33495, 235.713) (2.34495, 
  236.44) (2.35495, 237.168) (2.36495, 237.896) (2.37495, 
  238.624) (2.38495, 239.352) (2.39495, 240.08) (2.40495, 
  240.809) (2.41495, 241.537) (2.42495, 242.265) (2.43495, 
  242.994) (2.44495, 243.723) (2.45495, 244.451) (2.46495, 
  245.18) (2.47495, 245.909) (2.48495, 246.638) (2.49495, 
  247.367) (2.50495, 248.096) (2.51495, 248.825) (2.52495, 
  249.554) (2.53495, 250.284) (2.54495, 251.013) (2.55495, 
  251.743) (2.56495, 252.472) (2.57495, 253.202) (2.58495, 
  253.932) (2.59495, 254.661) (2.60495, 255.391) (2.61495, 
  256.121) (2.62495, 256.851) (2.63495, 257.581) (2.64495, 
  258.311) (2.65495, 259.042) (2.66495, 259.772)
  };
  \addplot[color=black,no markers,style={ultra thick}] coordinates{
(2.66495, 259.772)(2.67495, 260.502)(2.68495, 261.233)(2.69495,
   261.964)(2.70495, 262.695)(2.71495, 263.426)(2.72495, 
  264.157)(2.73495, 264.888)(2.74495, 265.62)(2.75495, 
  266.351)(2.76495, 267.083)(2.77495, 267.815)(2.78495, 
  268.547)(2.79495, 269.28)(2.80495, 270.012)(2.81495, 
  270.745)(2.82495, 271.478)(2.83495, 272.211)(2.84495, 
  272.945)(2.85495, 273.678)(2.86495, 274.412)(2.87495, 
  275.146)(2.88495, 275.881)(2.89495, 276.615)(2.90495, 
  277.35)(2.91495, 278.085)(2.92495, 278.821)(2.93495, 
  279.556)(2.94495, 280.292)(2.95495, 281.028)(2.96495, 
  281.764)(2.97495, 282.501)(2.98495, 283.238)(2.99495, 
  283.975)(3.00495, 284.713)(3.01495, 285.451)(3.02495, 
  286.189)(3.03495, 286.927)(3.04495, 287.666)(3.05495, 
  288.405)
  };
  \end{axis}
    \node[rotate=26] at (2.5,3.6) {$\dot \Sigma_{\textrm{dff}}$};
  \node[rotate=-26] at (2.5,1.2) {$\dot \Sigma_\mathrm{rct}-\dot \Sigma^{\rm h}$};
\end{tikzpicture}
\caption{
	Analytical result for the EPR of reaction $\dot \Sigma_{\mathrm{rct}}$ (dashed) and diffusion $\dot \Sigma_{\mathrm{dff}}$ (solid) in the stable stationary state as function of the concentration $y_2$.
	The EPR of the (unstable for $y_2>y_2^{\textrm{c}}$) homogeneous state  $\dot \Sigma^{\rm h}$ is subtracted from the former to show the effect of the pattern formation, i.e. decreasing reaction dissipation at the expense of diffusion dissipation.
	\emph{Inset:} The total entropy production shows no singularity at the phase transition.
	All results correspond to the weakly reversible case $k_{-\rho}=10^{-4}  \ll k_{+\rho}=1$, and $y_3=y_4=10^{-4} $, $y_1=2$, $D_{x_1}=1$, $D_{x_2}=10$.
}
\label{fig:EPR}
\end{figure}
%
The critical values $y_2^\mathrm{c}$ and $q_\mathrm{c}$ are determined by the condition of marginal stability of the homogeneous state: they are the smaller values for which the matrix $\partial_x \caJ{(x_1^\mathrm{h},x_2^\mathrm{h})}$ (evolving linearized perturbations) acquires a zero eigenvalue, the corresponding eigenvector being $(1,u_{x_2})^\mathsf{T}$. 
Near the onset of instability one can treat ${\epsilon=(y_2-y_2^\mathrm{c})/y_2^\mathrm{c}} \ll 1$ as a small parameter and carry out a perturbation expansion in powers of $\epsilon$. This leads to the  amplitude equation
 for  $A(r,t)$  \cite{pena02},
 \begin{align}
 \label{amp}
\tau \partial_t A= \epsilon A - \alpha |A|^2 A + \xi \partial_r^2 A,
\end{align}
which  describes an exponential growth from an initial small perturbation $A(r,0) \simeq 0$ followed by a late-time saturation due to the nonlinear terms in \eqref{brus}. Amplitude equations provide a general quantitative description of pattern formation in several systems near the onset of instability \cite{cross09}, irrespective of the details of the underlying physical process that are subsumed into the effective coefficients $\tau$, $\alpha$, and $\xi$.
Since \eqref{amp} can be seen as a gradient flow in a complex Ginzburg--Landau potential, pattern formation is usually considered a \emph{dynamical} phase transition \cite{aranson02}.
Here, using an analytical approximate solution to \eqref{brus} valid for $\epsilon \ll 1$, we show that the phenomenon is in fact a genuine \emph{thermodynamic} phase transition identified by the appearance of a kink singularity at $y_2^\mathrm{c}$ in the  nonequilibrium free energy $\caG(y_2)$. 
The semigrand canonical free energy of Fig.~\ref{fig:G} is calculated taking the stationary stable solution corresponding to a given value of $y_2$, i.e. the homogenous one for $y_2 \leqslant y_2^\mathrm{c}$ and the patterned one for $y_2 > y_2^\mathrm{c}$, namely
\begin{equation}
\begin{pmatrix}
x_1^\mathrm{p}(r)\\
x_2^\mathrm{p}(r)
\end{pmatrix}
\sim
\begin{pmatrix}
1 \\
u_\mathrm{x_2}
\end{pmatrix}
 \sqrt{\frac{\epsilon}{\alpha}}
 2\cos(q_c r)
  \end{equation}
 The physical meaning of the kink at $y_2=y_2^\mathrm{c}$ is best understood noticing that the quantity $\partial \caG /\partial \mu_{Y_2}= \dot W_{\mathrm{driv}}/\dt \mu_{Y_2}$ is the driving work upon a quasi-static manipulation of the chemical potential $\mu_{Y_2}$.  Interestingly, the total EPR shows no singularity at the transition (cf. Fig.~\ref{fig:EPR}): moving across $y_2^\mathrm{c}$ the EPR of reaction $\dot \Sigma_{\mathrm{rct}}$ decreases with respect to the homogeneous state value $\dot \Sigma^\mathrm{h}$ while a non zero EPR of diffusion appears, their sum being continuous. 
 
\emph{Conclusion---}We presented the nonequilibrium thermodynamics of RDS and exemplified the theory with the application to the Brusselator model.
We went beyond the conventional treatment of classical nonequilibrium thermodynamics \cite{kjelstrup08} in two respects: avoiding to linearize the chemistry, i.e. to oversimplify reaction affinities to currents times Onsager coefficients; explicitly  building thermodynamic potentials that act as Lyapunov functions in the relaxation to equilibrium, provide minimum work principles, and reveal the existence of nonequilibrium phase transitions. The framework paves the way to study the energy cost of pattern manipulation and information transmission in complex chemical systems \cite{zadorin15, epstein2016,halatek18}.
 
We acknowledge funding from the National Research Fund of Luxembourg (AFR PhD Grant 2014-2, No.~9114110) and the European Research Council project NanoThermo (ERC-2015-CoG Agreement No.~681456). 

\bibliography{NotesTuring}

\end{document}